\documentclass[]{llncs}
\usepackage{epic,epsf,amsmath,latexsym,amssymb,ifthen,color}
\usepackage{amsmath,latexsym,amssymb}
\usepackage{times,pstricks,graphicx,rotating,hhline}
\usepackage{setspace}

\newcommand{\toto}{xxx}
\newenvironment{proofT}{\noindent{\bf
Proof }} {\hspace*{\fill}$\Box_{Theorem~\ref{\toto}}$\par\vspace{3mm}}
\newenvironment{proofL}{\noindent{\bf
Proof }} {\hspace*{\fill}$\Box_{Lemma~\ref{\toto}}$\par\vspace{3mm}}
\begin{document}

\newcommand{\remove}[1]{}



\title{{\bf Core Persistence in Peer-to-Peer Systems:\\
            Relating Size to Lifetime}}

\author{Vincent Gramoli~~ Anne-Marie Kermarrec~~ 
        Achour Mostefaoui~~  \\ Michel  Raynal~~  
        Bruno  Sericola  \\~\\ 
        IRISA (Universit\'e de Rennes 1 and INRIA)\\
       Campus de Beaulieu, 35042 Rennes,  France \\ 
}

\author{Vincent Gramoli \and Anne-Marie Kermarrec\\ 
        Achour Mostefaoui \and Michel  Raynal \and 
        Bruno  Sericola}

\institute{IRISA (Universit\'e de Rennes 1 and INRIA)\\
       Campus de Beaulieu, 35042 Rennes,  France.\\
			 \email{\{vgramoli,akermarr,achour,raynal,bsericol\}@irisa.fr}
       }
       
\authorrunning{Vincent Gramoli et al.}       

\tocauthor{
	Vincent Gramoli (IRISA, Universit\'e de Rennes 1)
	Anne-Marie Kermarrec (IRISA, INRIA)
	Achour Mostefaoui (IRISA, Universit\'e de Rennes 1)
	Michel Raynal (IRISA, Universit\'e de Rennes 1)
	Bruno Sericola (IRISA, INRIA)
}

\date{}


\maketitle


\begin{abstract}
Distributed systems are now both very large and highly dynamic.
Peer to peer overlay networks have been proved efficient to cope with this new deal 
that  traditional approaches  can no  longer accommodate.  While the challenge of 
organizing
peers in an overlay network has generated a lot of interest leading to a
large number  of solutions,  maintaining  critical data  in such  a network
remains an open issue.  In this  paper, we are interested  in defining the
portion of nodes and frequency  one  has to probe, given the churn observed
in the system, in order to achieve a given  
probability of maintaining the persistence of some critical data. 
More  specifically, we provide  a clear  result relating  the size  and the
frequency of the probing set along with its proof as well as an analysis of
the  way  of leveraging   such  an information  in  a  large scale  dynamic
distributed system.

~   
\end{abstract}


\section{Introduction}

\paragraph{Context of the paper.}
Persistence of critical data in distributed applications is a crucial problem.
Although static systems have experienced many solutions, 
mostly relying
on defining the right degree  of replication, this remains  an open
issue in the context of dynamic systems. 

Recently, peer to peer (P2P) systems became popular 
as they
have been proved efficient to cope with the scale shift observed in
distributed systems.  A P2P system is a dynamic system that allow
peers (nodes) to join or leave the system. 
In the meantime, a natural tendency to trade 
strong deterministic guarantees for probabilistic ones aimed
at coping with both scale and dynamism.
Yet, quantifying bounds of guarantee that can be achieved probabilistically is 
very important for the deployment of applications. 

More  specifically, a typical issue is to ensure that despite dynamism 
some critical data is not lost.
The set of nodes owning a copy of the critical data is called a {\it core}
(distinct cores can possibly co-exist, each associated with a particular
data).   

Provided that core nodes remain long enough in the system, 
a ``data/state  transfer'' protocol can transmit the critical data from nodes to nodes.
This ensures that a new core of nodes in the system will 
keep track of the data.
Hence, such protocols provide data persistence despite the uncertainty of the system state
involved by the dynamic evolution of its members. 

There is however an inherent tradeoff in the use of such protocols.  
If the policy that is used is too  conservative, 
the data transfer protocol might be executed too often, thereby consuming
resources and increasing the whole system overhead. 
Conversely, if the protocol is executed too rarely, 
all nodes owning a copy of the data may leave (or crash) before  
a new protocol execution, and the data would be lost.
This fundamental tradeoff is the main problem addressed in 
this paper.

\paragraph{Content of the paper.}
Considering the previous context, we are interested in providing some 
probabilistic guarantees of maintaining a \textit{core} in the system.
More precisely, given the churn observed in the system, we aim at 
maintaining the persistence of some critical data.
To this end, we are interested in defining the portion of nodes 
that must be probed, as well as the frequency to which this probe must 
occur to achieve this result with a given probability.
This boils down to relating the size and the frequency of the probing set
according to a target probability of success and the churn observed in the
system. 

The investigation of the previous tradeoff relies on  critical
parameters. One of them is naturally the size of the core. 
Two other parameters are the percentage of nodes that enter/leave the
system per time unit, 
and the duration 
during which we observe the system.  We first assume that, per time unit,
the number of entering nodes is the same as the number of leaving
nodes. 
In other words, the number of nodes remains constant.

Let $S$ be the system at some time $\tau$. It is composed of 
$n$ nodes including a subset of $q$ nodes defining a core $Q$ for a 
given critical data.  Let $S'$ be the system at time $\tau+\delta$.
Because of the system evolution, 
some nodes owning a copy of the  critical data at time $\tau$ might have
left the system at time $\tau+\delta$ (those nodes are in $S$ and not in
$S'$).  So, an important question is the following: 
``Given a set  $Q'$ of $q'$  nodes of $S'$,  what is the probability that
$Q$ and $Q'$ do intersect?'' 
We derive an explicit expression of this probability as a function of
the parameters characterizing the dynamic system.
This allows us to  compute some  of them
when other ones are fixed. 
This provides distributed 
applications with the opportunity to set a tradeoff between a probabilistic 
guarantee of achieving a core  and the overhead involved computed either 
as the number of nodes probed or the frequency at which the probing set needs
to be refreshed.

\paragraph{Related work.}
As mentioned above, P2P systems have received a great deal of
attention both in academia  
and  industry  for  the  past  five  years. More  specifically,  a  lot  of
approaches have been proposed to  
create whether they are structured,  such as Chord \cite{stoica.i2001}, CAN
\cite{ratnasamy.s2001}  or Pastry  \cite{rowstron.a2001c},  or unstructured
\cite{lpbcast,ganesh.a2003,JGKvS04}. Maintenance of such overlay networks
in the presence of high churns has also been studied as one
of the  major goal of P2P overlay networks~\cite{LBK02}.  The parameters 
impacting on connectivity and routing capabilities in P2P overlay networks  
are now well understood. 

In structured  P2P networks, routing tables contain critical information
and refreshment must occur with some frequency depending on the churn 
observed in the  network~\cite{castro2004} to achieve routing 
capabilities.
%
%
For instance in Pastry, the  size of the leaf set (set of nodes whose 
identities are numerically the closest to current node identity) and its 
maintenance protocol can be tuned to achieve the routing within reasonable  
delay stretch and low overhead.
%
%
%
Finally, there has been approaches evaluating the number of locations
to which a data has to be replicated
in the  system in  order to be  successfully searched by  flooding-based or
random walk-based algorithms~\cite{cohen2002}.  These approaches do 
not consider specifically churn in their analysis. 
In this paper churn is a primary concern.  The result of this work can
be applied to any P2P network, regardless of its structure, in order
to maintain critical data by refreshment at sufficiently many locations.
%

The use of a base core to extend protocols designed for static systems to 
dynamic systems has been investigated in \cite{MRT05}. 
Persistent cores share some features with quorums (i.e., mutually 
intersecting sets). Quorums originated a  long time 
ago  with   majority   voting  systems \cite{G79,T79}  introduced to ensure
data   consistency.   
More recently, quorum reconfiguration~\cite{LS02,CGGMS05} 
have been proposed to face system dynamism while 
guaranteeing atomic consistency: this application outlines the strength of
such dynamic quorums.
Quorum-based  protocols for searching  objects in  P2P
systems  are proposed in~\cite{MTK06}.  Probabilistic  quorum  systems  
have been introduced in~\cite{MRWW01}.  They use randomization to
relax the strict intersection property to a probabilistic one.  They have
been extended to dynamic systems in~\cite{AM05}.

\paragraph{Roadmap.}
The paper is organized as follows. 
Section~\ref{sec:model}
defines the system model. Section~\ref{sec:analysis} describes
our dynamic system analysis and our probabilistic results. 
Section~\ref{sec:interpretation} interprets the
previous formulas and shows how to use them  to  control
the uncertainty of the key parameters of P2P applications.  
Finally, Section \ref{sec:discussion} 
concludes the paper.

\section{System model}
\label{sec:model}
The system model, sketched in the introduction is simple. 
The system consists of $n$ nodes. It is dynamic in the following sense. 
For the sake of simplicity, let $n$ be the size of the system.
Every time unit, $cn$  nodes leave the system and $cn$ 
nodes enter the system, where $c$ is the percentage of nodes that enter/leave
the system per time unit; this  can be seen as  new nodes ``replacing''
leaving nodes.  Although monitoring the leave and join rates of a large-scale 
dynamic system remains an open issue, it is reasonable to assume join and leave
are tightly correlated in P2P systems. 
A more realistic model would take in account variation of the system size 
depending for instance, on night-time and day-time as observed in~\cite{SGG02}.


A node leaves the system either voluntarily or because it crashes.  
A node that leaves the system  does not enter it later.  (Practically, 
this means that, to re-enter the system, a node that has left 
must be considered as a new node; all its previous knowledge of the system 
state is lost.) 
For instance, initially (at time $\tau$), assume 
there  are $n$ nodes  (identified  from  $1$ to
$n$; let us take $n=5$ to  simplify).  Let $c=0.2$, which means that, 
every time unit,  $nc=1$  node changes (a node disappears and a new 
node replaces it).  
That is, at time $\tau+1$, one node leaves the system and another one
joins.  From now on, observe that next leaving nodes are either
nodes that were initially in the system or nodes that joined
after time $\tau$.
%
%


\section{Relating the key parameters of the dynamic system}
\label{sec:analysis}

This   section answers the  question posed  in the  introduction, namely,
given a  set $Q(\tau)$ of nodes  at time $\tau$  (the core), 
and  a set  $Q(\tau')$ of  nodes at   time  $\tau'=\tau+\delta$, what
is the  probability  of the event ``$Q(\tau)\cap Q(\tau')\neq\emptyset$''. 
In the remaining of this paper, we assume that both $Q(\tau)$ and $Q(\tau')$ 
contain $q$ nodes, since an interesting goal is to minimize both the 
number of nodes where the data is replicated and the number of nodes one has 
to probe to find the data.
%
Let an  {\it initial} node  be a  node that belongs  to the system  at time
$\tau$.  Moreover,  without  loss   of  generality,  let  $\tau=0$  (hence,
$\tau'=\delta$). 

\begin{lemma}
\label{lemme-valeur-de-C}
Let $C$ be the ratio of initial nodes that are replaced after
$\delta$ time units. We have $C=1-(1-c)^\delta$. 
\end{lemma}

\begin{proofL}
We claim that the number of initial nodes that are still in the system
after $\delta$ time units is $n(1-c)^{\delta}$. 
The proof is by  induction on the time instants. Let us  remind that $c$ is
the percentage of nodes that are replaced in one time unit. 
For the Base case, at  time $1$,  $n-n c=n(1-c)$  nodes have  not been
replaced. 
For the induction case, let us assume  that at time $\delta-1$, the number of
initial nodes  that have  not been replaced is $n(1-c)^{\delta-1}$. 
Let us  consider the  time instant  $\delta$.  
The number of initial nodes that are not replaced after $\delta$ time units
is  $n(1-c)^{\delta-1}  - n(1-c)^{\delta-1}c$, i.e.,  $n(1-c)^{\delta}$,
which proves the claim. 
It follows from the previous claim that the number of initial nodes that are 
replaced during $\delta$ time  units is $n-n(1-c)^\delta$. 
Hence, $C= (n-n(1-c)^\delta)/n= 1-(1-c)^\delta$. 
\renewcommand{\toto}{lemme-valeur-de-C}
\end{proofL}

Given a  core of $q$ nodes at  time $\tau$ (each having a copy of the
critical  data), 
the following theorem  gives the probability that, at time $\tau'=\tau+\delta$, an 
arbitrary node cannot obtain the data when it queries $q$ nodes arbitrarily 
chosen.  

For this purpose, using result of Lemma~\ref{lemme-valeur-de-C} we take the 
number of elements that have left the system during the period $\delta$ as 
$\alpha = \lceil Cn\rceil = \lceil(1 - (1 - c)^\delta) n\rceil$.
This number 
allows us to evaluate the aforementioned probability.
%

\begin{theorem}
	\label{theo-proba-pas-vert}
	Let $x_1, ..., x_q$ be any node in the system at time $\tau'=\tau +  \delta$. The
	probability that none of these nodes belong to the initial core is
	$$\frac{\sum^{b}_{k=a}
	\left[\displaystyle{{n+k-q \choose q}{q \choose k}{n-q \choose \alpha-k}}\right]}{\displaystyle{n \choose q}\displaystyle{n \choose \alpha}},$$
	where 
	$\alpha = \lceil(1 - (1 - c)^\delta) n\rceil$, 
	$a = \max(0, \alpha - n + q)$, and 
	$b = \min(\alpha, q)$.
\end{theorem}

\begin{proofT}
The problem we have to solve can be represented in the following way:

	The system is an urn containing  $n$  balls (nodes), such that, 
	initially, $q$ balls are green (they represent the initial core $Q(\tau)$
	and are represented by the set $\cal Q$ in Figure \ref{fig-proba}),  
	while the $n-q$ remaining balls are black. 
	 
	We randomly draw $\alpha=\lceil Cn \rceil$ balls from the urn (according to a uniform 
	distribution), and paint them  red. These $\alpha$ balls represent the 
	initial nodes that are replaced by new nodes after $\delta$  units of time 
	(each of these balls was  initially  green or black). 
	After it has been  colored red, each of these balls is put back 
	in the urn (so, the urn contains again $n$ balls).  
	
	We then obtain the system as described in the right part of Figure
	\ref{fig-proba}    (which   represents   the    system   state    at   time
	$\tau'=\tau+\delta$).  The set $\cal A$ is
	the set of balls that have been painted red.  $\cal Q'$ is the
	core set $\cal Q$ after  some of its balls have been painted  red
	(these  balls  represent the nodes  of the core  that have  left  the
	system). 
	This means the set ${\cal Q}' \setminus {\cal A}$, that we denote by ${\cal E}$, 
	contains all the green balls and only them.
	
	We denote by $\beta$ the number of balls in the set ${\cal Q'} \cap {\cal A}$.
	It is well-known that 
	$\beta$ 
	has a hypergeometric distribution, i.e., 
	for $a\leq k\leq b$ where $a=\max(0,\alpha-n+q)$ and $b=\min(\alpha,q)$,
	we have

	\begin{eqnarray}
	\Pr[\beta=k] &=& \frac{\displaystyle{{q \choose k}{n-q \choose \alpha-k}}}{\displaystyle{{n \choose \alpha}}}. \label{eq:hyperg}
	\end{eqnarray}

	We finally draw randomly and successively $q$ balls $x_1, ..., x_q$ from the urn (system at time $\tau'$)
	without replacing them.
	The problem consists in computing the probability of the event
	\{none of the selected balls $x_1, ..., x_q$ are green\}, which can be written as 
	$\Pr[x_1\notin {\cal E}, ..., x_q\notin {\cal E}]$.

\begin{figure}[h]
\begin{center}
\scalebox{0.2}[0.2]
{\input{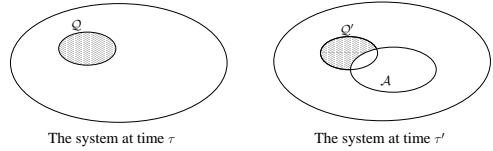}}
\end{center}
\caption{The system  at times $\tau$ and $\tau'=\tau+\delta$}
\label{fig-proba} 
\vspace{-0.5cm}
\end{figure} 

As $\{x\in {\cal E}\} \Leftrightarrow 
\{x\in {\cal  Q}'\} \cap \{x\notin  {\cal  Q}'\cap {\cal A}\}$,  we have (taking
the contrapositive) $\{x\notin {\cal E}\} \Leftrightarrow 
\{x\notin {\cal Q}' \} \cup \{ x\in  {\cal Q}'\cap {\cal A}\}$, from which we can
conclude
%
$\Pr[x\notin {\cal E}]=\Pr[\{x\notin  {\cal Q}'\}~\cup~ \{x\in  {\cal Q}'\cap {\cal A}\}].$
As the events \{$x\notin {\cal Q}'$\} and 
 \{$x\in  {\cal Q}'\cap {\cal A}$\} are disjoints, we obtain
%
$ \Pr[x\notin {\cal E}]=
   \Pr[x\notin {\cal Q}'] + 
   \Pr[x\in  {\cal Q}'\cap {\cal A}].$
The system contains $n$ balls. The number of balls in ${\cal Q'}$, ${\cal A}$ and
${\cal Q'} \cap {\cal A}$ is equal to $q, \alpha$ and $\beta$, respectively.  
%
%
%
%
Since there is no replacement, we get, 
%
%

	{\small	
\begin{eqnarray}
	\Pr[x_1 \notin {\cal E}, ..., x_q \notin {\cal E} ~\big/ ~ \beta = k] &= 					
	  \sum^{b}_{k=a}\prod^{q}_{i=1}\left(1-\frac{q-k}{n-i+1}\right) = \sum^{b}_{k=a} \frac{\displaystyle{n-q+k \choose q}}{\displaystyle{n \choose q}}. \label{eq:cond}
\end{eqnarray}}
%
To uncondition the aforementioned result~(\ref{eq:cond}), we simply multiply it by~(\ref{eq:hyperg}), leading to
\begin{eqnarray}
	  \Pr[x_1 \notin {\cal E}, ..., x_q \notin {\cal E}] 
	  &=& \frac{\sum^{b}_{k=a}\left[\displaystyle{n+k-q \choose q}{q \choose k}{n-q \choose \alpha-k}\right]}{\displaystyle{n \choose q}\displaystyle{n \choose \alpha}}. \notag
\end{eqnarray}
\remove{

Since there is no replacement, we clearly have, for any $i$, $1\leq i \leq q$ 

\begin{eqnarray}
  \Pr[x_i\notin {\cal E} ~\big/ ~ \beta = k] &=&
   \left(1-\frac{q}{n-i+1}\right) + \frac{k}{n-i+1} = \left(1-\frac{q-k}{n-i+1}\right). \notag
\end{eqnarray}

The probability that none of the selected balls are green is then:

\begin{eqnarray}
	  \Pr[x_1 \notin {\cal E}, ..., x_q \notin {\cal E}] &=& 
	  \sum^{b}_{k=a} \Pr[x_1 \notin {\cal E}, ..., x_q \notin {\cal E}  ~\big/ ~ \beta=k] \times \Pr[\beta=k]. \notag
\end{eqnarray}
Fixing $\beta$ makes the events $\forall i$, $1\leq i \leq q$ \{$x_i \notin {\cal E}$\} independent.  Thus we obtain
\begin{eqnarray}
	  \Pr[x_1 \notin {\cal E}, ..., x_q \notin {\cal E}] &=& \sum^{b}_{k=a}\left(1-\frac{q-k}{n}\right)\left(1-\frac{q-k}{n-1}\right)...\left(1-\frac{q-k}{n-q+1}\right)\Pr[\beta=k], \notag \\ 
	  &=& \sum^{b}_{k=a}\prod^{q-1}_{i=0}\left(1-\frac{q-k}{n-i}\right)\Pr[\beta=k]. \label{eq:sum}
\end{eqnarray}

Replacing probability (\ref{eq:hyperg}) in (\ref{eq:sum}) leads to the following result:
	
	\begin{eqnarray}
	  \Pr[x_1 \notin {\cal E}, ..., x_q \notin {\cal E}] &=& \sum^{b}_{k=a}\prod^{q-1}_{i=0}\left(1-\frac{q-k}{n-i}\right) \frac{\displaystyle{{q \choose k}{n-q \choose \alpha-k}}}{\displaystyle{n \choose \alpha}},\notag \\
	  &=& \frac{1}{\displaystyle{n \choose \alpha}}\sum^{b}_{k=a}\left[\frac{(n+k-q)(n-q)!}{(n+k-2q)!n!} \displaystyle{q \choose k}{n-q \choose \alpha-k}\right], \notag \\
	  &=& \frac{1}{\displaystyle{n \choose \alpha}}\sum^{b}_{k=a}\left[\frac{\displaystyle{n+k-q \choose q}}{\displaystyle{n \choose q}} {q \choose k}{n-q \choose \alpha-k}\right], \notag \\
	  &=& \frac{1}{\displaystyle{n \choose \alpha}\displaystyle{n \choose q}}\sum^{b}_{k=a}\left[\displaystyle{n+k-q \choose q}{q \choose k}{n-q \choose \alpha-k}\right]. \notag
\end{eqnarray}
	
	} 
\renewcommand{\toto}{theo-proba-pas-vert}\end{proofT}
\section{From formulas to parameter tuning}
\label{sec:interpretation}

In the  previous section, we  have provided a  set of formulas that  can be
leveraged   and    exploited   by   distributed    applications   in   many
ways. Typically,  in a  P2P system,  the churn rate  is not  negotiated but
observed\footnote{Monitoring the churn  rate of  a system,  although very
interesting, is   out  of  the   scope  of  this   paper.}.  Nevertheless,
applications deployed on P2P overlays  
 may   need  to  choose   the  probabilistic   guarantees  
 that a node of the initial core is probed.
 Given such a probability, the application 
may fix  either  the  size of the probing set of  nodes or  the
frequency at  which the core  needs  to be re-established from the current
set of nodes (with the  help  of an appropriate data transfer protocol).

This section  exploits  the previous  formula  to relate  these
various elements. More precisely, we 
provide the various  relations existing between the three  factors that can
be tuned by an application designer: 
the size of the probing set  $q$, the frequency of the probing $\delta$, and
the probability of achieving a core characterized by $p= 1-\epsilon$.
(For the sake of clarity all along this section, a ratio $C$, or  $c$,
is sometimes expressed as a  percentage.  
Floating point numbers on the $y$-axis are represented in 
their mantissa and exponent numbers.)

\paragraph{Relation linking $c$ and $\delta$.}

The first parameter that we formalized is $C$, that can be interpreted as
 the rate of dynamism in the system.  
$C$ depends  both  on the churn rate  ($c$) observed in the  system and the
 probing frequency  ($1/\delta$).
More specifically, we foresee here a scenario in which an application designer
would consider tolerating a churn $C$ in order to define 
the  size of   a core  and  thus ensure  the persistence  of some  critical
data. For example, an application may need to tolerate a churn rate of 10\%
in the system, meaning that the persistence of some critical data should be
ensured as  long as  up to  10\% of the  nodes in  the system  change  over
time.  Therefore, depending  on the  churn  observed and  monitored in  the
system, we are able to  define  the longest period $\delta$ before which the
core  should be re-instantiated on a set of the current nodes.  One of  the
main   interest of  linking  $c$   and  $\delta$  is  that if   $c$ varies
over time,  $\delta$  can  be  adapted   accordingly without  compromising
the  initial requirements of the application.  

\begin{figure}
\vspace{-0.5cm}
\begin{center}
	\setlength{\psunit}{0.55cm}
	\psset{unit=\psunit}
	\begin{pspicture}(0,0)(11,11)
	\rput(5.5,5.5){\includegraphics[width=10\psunit, height=10\psunit]{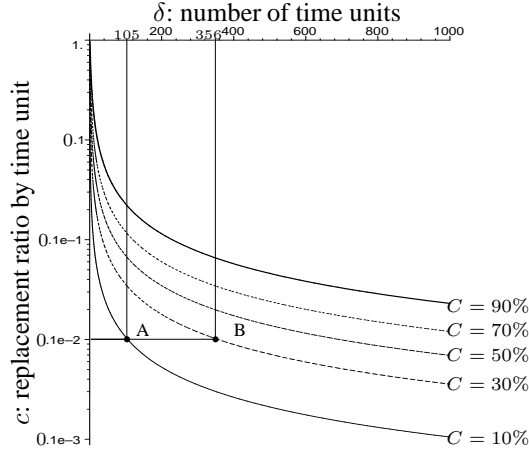}}
	\rput(0,1){\begin{rotate}{90}$c$: replacement ratio by time unit\end{rotate}}
	\rput(6,11){$\delta$: number of time units}
	\rput(11.1,0.7){\scriptsize $C=10\%$}
	\rput(11.1,2){\scriptsize $C=30\%$}
	\rput(11.1,2.7){\scriptsize $C=50\%$}
	\rput(11.1,3.3){\scriptsize $C=70\%$}
	\rput(11.1,3.9){\scriptsize $C=90\%$}
	\rput(2.37,3.07){\huge .}\rput(2.75,3.3){\scriptsize A}
	\psline[linewidth=0.1pt]{-}(1.5,3.07)(2.37,3.07)
	\psline[linewidth=0.1pt]{-}(2.37,10.35)(2.37,3.07)
	\rput(4.51,3.07){\huge .}\rput(5.1,3.3){\scriptsize B}
	\psline[linewidth=0.1pt]{-}(1.5,3.07)(4.51,3.07)
	\psline[linewidth=0.1pt]{-}(4.51,10.35)(4.51,3.07)
	\rput(0.95,7.88){\scriptsize $0$}
	\rput(0.35,5.48){\scriptsize $0$}
	\rput(0.30,3.05){\scriptsize $0$}
	\rput(0.30,0.65){\scriptsize $0$}
	\rput(2.37,10.43){\tiny $105$}
	\rput(4.35,10.43){\tiny $356$}
	\end{pspicture}
\end{center}
\caption{Evolution of the pair $(c,\delta)$ for given values of $C$}
\label{courbes-c-et-delta}
\vspace{-0.5cm}
\end{figure}

More formally, Lemma   \ref{lemme-valeur-de-C}  provides  an explicit value
of   $C$  (the  ratio   of  initial   nodes   that   are  replaced)   as  a
function  of $c$   (the replacement ratio per time  unit) and $\delta$ (the
number  of time  units).  Figure  \ref{courbes-c-et-delta}  represents this
function  for several  values of  $C$. More  explicitly, it   depicts  on a
logarithmic  scale the curve $c=1-\sqrt[\delta]{1-C}$ (or equivalently, the
curve  $\delta= \frac{\log (1-C)}{\log(1-c)}$).  
As an example, the  curve associated with $C=10\%$  indicates that $10\%$ 
of the initial nodes have been replaced after  $\delta=105$ time
units (point A, Figure \ref{courbes-c-et-delta}), when the replacement ratio 
is $c=10^{-3}$ per time unit. 
Similarly, the same replacement ratio per time unit entails the replacement 
of  $30\%$  of  the  initial  nodes  when  the  duration  we  consider  is
$\delta=356$  time  units (point B, Figure \ref{courbes-c-et-delta}).  
The system designer can benefit from these values to 
better appreciate the way the system evolves according to the assumed
replacement ratio per time unit.
To summarize, this result can be used as follows. In a system, 
aiming at tolerating a churn of $X$\% of the nodes, our goal is to
provide an application with  the corresponding value of $\delta$, 
knowing the churn   $c$ observed in the system.  This gives the opportunity
to adjust  $\delta$ if   $c$ changes over time.  

\paragraph{Relation linking  the core size $q$ and $\epsilon$}

Now, given  a value $C$ set by  an application developer, there are still 
two parameters that may influence either the overhead of maintaining a core
in the  system, or the  probabilistic guarantee of  having such a  core. The
overhead may be measured in a straightforward manner in this context as the
number of  nodes that  need to  be probed, namely  $q$. Intuitively,  for a
given  $C$, as  $q$ increases,  the probability  of probing  a node  of the
initial core  increases. In this section, we define how much 
these parameters are related.

\begin{figure}\vspace{-0.5cm}
	\begin{center}
		\setlength{\psunit}{0.55cm}
		\psset{unit=\psunit}

		\begin{pspicture}(0,0)(21,11)
		\rput(5.02,5.75){\includegraphics[width=10\psunit, height=10\psunit]{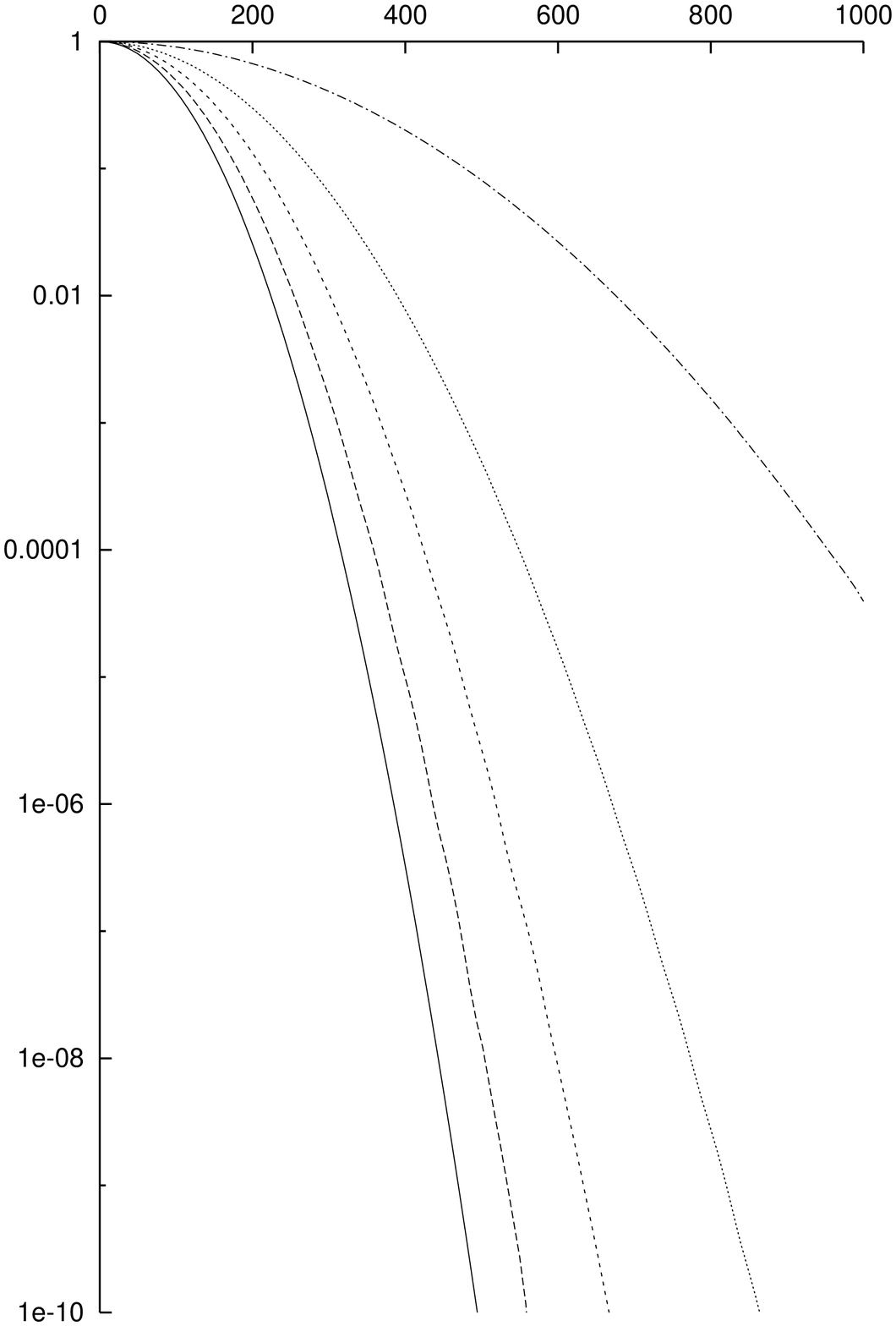}}
		\rput(0,2){\begin{rotate}{90}$\epsilon$: the probability $1-p$\end{rotate}}
		\rput(6,11){$q$: the core size}
		\rput(2.7,5){\tiny $C=10\%$}
		\psline{->}(3.6,5)(4.4,5)
		\rput(2.7,5.5){\tiny $C=30\%$}
		\psline{->}(3.6,5.5)(4.7,5.5)
		\rput(2.7,6){\tiny $C=50\%$}
		\psline{->}(3.6,6)(5.05,6)
		\rput(2.7,6.5){\tiny $C=70\%$}
		\psline{->}(3.6,6.5)(5.9,6.5)
		\rput(2.7,7){\tiny $C=90\%$}
		\psline{->}(3.6,7)(8.7,7)
		\rput(3.68,7.53){\huge .}\rput(3.45,7.7){\scriptsize A}
		\psline[linewidth=0.1pt]{-}(1.5,7.53)(3.68,7.53)
		\psline[linewidth=0.1pt]{-}(3.68,10.35)(3.68,7.53)
		\rput(4.48,7.53){\huge .}\rput(4.22,7.7){\scriptsize B}
		\psline[linewidth=0.1pt]{-}(1.5,7.53)(4.48,7.53)
		\psline[linewidth=0.1pt]{-}(4.48,10.35)(4.48,7.53)
		\rput(3.8,10.2){\tiny 274}
		\rput(4.65,10.2){\tiny 369}
		\rput(5.5,-0.55){(a)}
		
		\rput(16.5,5.5){\includegraphics[width=10\psunit, height=10\psunit]{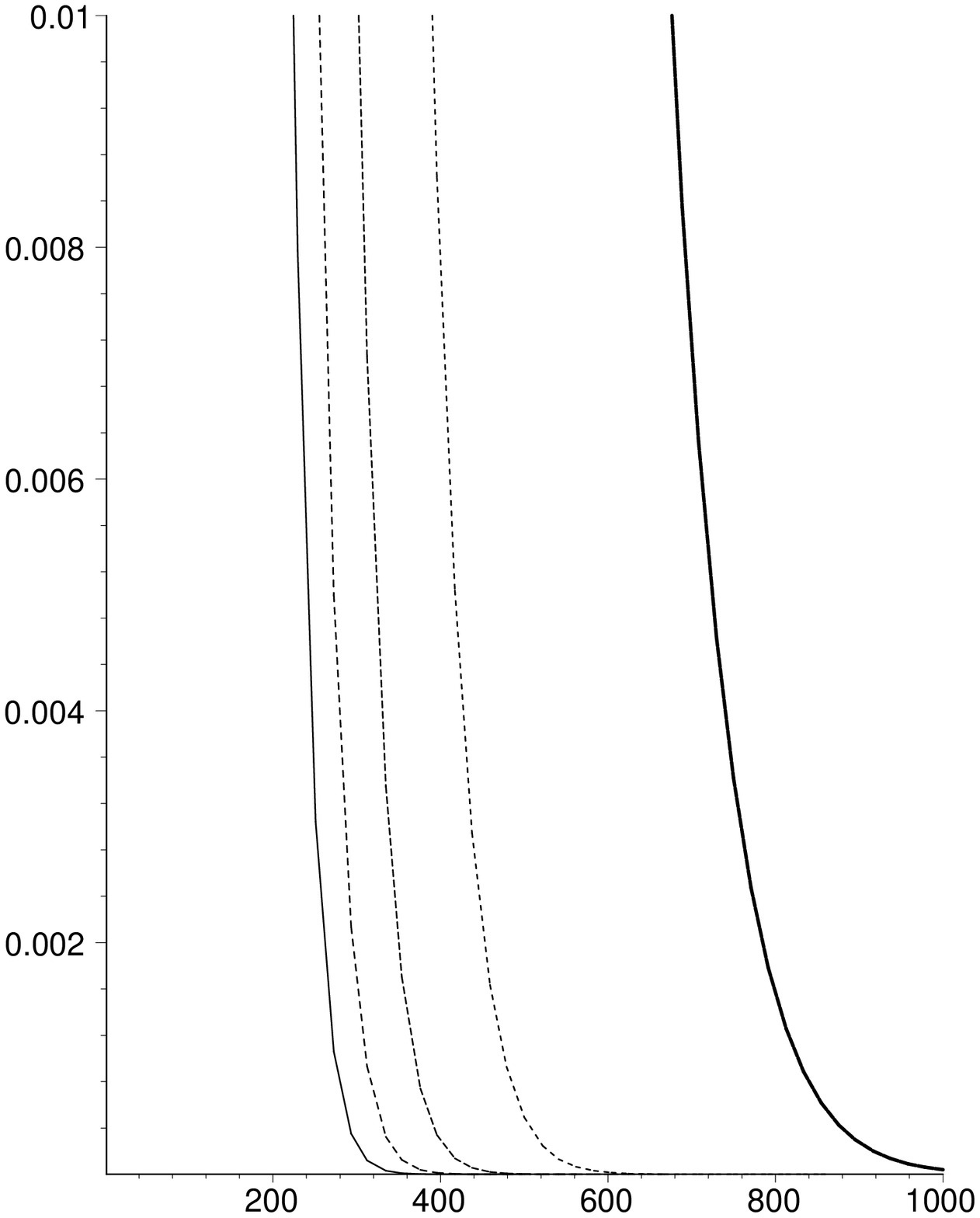}}
		\rput(11,2){\begin{rotate}{90}$\epsilon$: the probability $1-p$\end{rotate}}
		\rput(16.5,0){$q$: the core size}
		\rput(13.4,5){\tiny $C=10\%$}
		\psline{->}(14.3,5)(14.6,5)
		\rput(13.4,5.5){\tiny $C=30\%$}
		\psline{->}(14.3,5.5)(14.8,5.5)
		\rput(13.4,6){\tiny $C=50\%$}
		\psline{->}(14.3,6)(15.2,6)
		\rput(13.4,6.5){\tiny $C=70\%$}
		\psline{->}(14.3,6.5)(16,6.5)
		\rput(13.4,7){\tiny $C=90\%$}
		\psline{->}(14.3,7)(18.6,7)
		\psline[linewidth=0.1pt]{-}(12.4,10.4)(14.4,10.4)
		\psline[linewidth=0.1pt]{-}(14.4,10.4)(14.4,0.4)
		\rput(14.3,0.3){\tiny 224}
		\psline[linewidth=0.1pt]{-}(12.4,1.8)(14.8,1.8)
		\psline[linewidth=0.1pt]{-}(14.8,1.8)(14.8,0.4)
		\rput(14.9,0.3){\tiny 274}
		\rput(16.5,-0.55){(b)}
		\end{pspicture}
	\end{center}
	\caption{Non-intersection probability over the core size}
	\label{courbes-eps-et-q}
	\vspace{-0.5cm}
\end{figure}

Let us consider the value $\epsilon$ determined by Theorem 
\ref{theo-proba-pas-vert}. That value can be interpreted the following way:
$p = 1-\epsilon$  is the probability 
that, at time $\tau'=\tau+\delta$, one of the $q$ queries 
issued (randomly) by a node hits a node of the core. 
An important question is  then the following: How are $\epsilon$ and $q$
related? Or equivalently, how increasing the size of $q$ enable to decrease
$\epsilon$?  
This  relation is depicted in  Figure \ref{courbes-eps-et-q}(a) where
several curves are represented for $n=10,000$ nodes.

Each curve corresponds to a percentage of the initial nodes that have 
been replaced. (As an example, the  curve $30\%$ corresponds  to the case
where  $C=30\%$ of  the initial nodes have left the system; 
the way $C$, $\delta$ and $c$ are related has been seen previously.) 
%
%
%
%
Let us consider $\epsilon=10^{-3}$. The curves show that $q=274$ is
a sufficient core  size for not bypassing that value  of $\epsilon$ when up
to $10\%$ of the nodes  are replaced (point A, Figure~\ref{courbes-eps-et-q}(a)). 
Differently,   $q=274$ is  not sufficient when up to  $50\%$  of the  nodes
are  replaced;   in that   case, the  size  $q=369$  is required  (point B,
Figure~\ref{courbes-eps-et-q}(a)).

The curves of both Figure \ref{courbes-c-et-delta} and Figure~\ref{courbes-eps-et-q}(a) 
provide the system designer with realistic hints
to set the value of $\delta$ (deadline before which a data transfer
protocol establishing a new core has to be executed). 
Figure~\ref{courbes-eps-et-q}(b) is  a zoom of Figure~\ref{courbes-eps-et-q}(a) focusing on the small values of  $\epsilon$.  
It shows that, when $10^{-3} \leq \epsilon \leq 10^{-2}$, the probability 
$p= 1-\epsilon$ increases very rapidly towards 1, though the size of the 
core increases only very slightly.   As an   example, let   us
consider  the  curve associated with $C=10\%$ in  
Figure \ref{courbes-eps-et-q}(b). It shows that a
core of $q=224$ nodes ensures an intersection probability~$=1-\epsilon=0.99$,
and  a   core   of  $q=274$  nodes  ensures   an  intersection  probability
$=1-\epsilon=0.999$.

%

Interestingly,  this phenomenon  is similar to the {\it birthday
paradox}\footnote{The  paradox  is  with  respect  to  intuition,  not  with
respect to logics.} \cite{I95} that can be roughly summarized as follows.  
How many persons must be present in a room for two of them
to have the same birthday with probability $p=1-\epsilon$? 
Actually, for  that probability to  be greater than $1/2$, it is sufficient
that the number of persons in the room  be equal (only) to  $23$! 
When, there are $50$ persons  in the  room, the probability becomes $97\%$,
and  increases to $99.9996\%$ for $100$ persons. 
In  our  case, we observe a similar phenomenon: 
the  probability $p=1-\epsilon$  increases  very rapidly despite the fact
that  the frequency of the core size $q$ increases slightly.

In our case, this means that  the system designer 
can choose to slightly increase the size of the probing set $q$ (and 
therefore  only slightly increase the associated overhead)
while significantly increasing the probability to access a node of the core.


\paragraph{Relation linking $q$ and $\delta$}

So far, we have considered that an application may need to fix $C$ and 
then define the size of the probing set to achieve a given probability $p$ of 
success. There is another remaining trade-off that an application designer
might want to decide upon:  trading the size of the probing set with the
probing frequency while fixing the probability $p=1-\epsilon$
of intersecting the initial core. This is precisely defined by relating 
$q$ to $\delta$ for a fixed $\epsilon$.

In the following we investigate the way the size and lifetime of the core
are related when the required intersection probability is 
$99\%$ or  $99.9\%$.  We chose these values to better illustrate our purpose,
as we believe they reflect what could be expected by an application
designer.  For both  probabilities we present two different figures summarizing
the required  values of  $q$. 

%

\begin{figure}[!ht]
 \centering
 \resizebox{2.8in}{!}{
  \begin{tabular}{|c|c|ccc|}
  \hline
   
	 Intersection & Churn &  & Core size & \\ 
	 probability & $C=1-(1-c)^\delta$ & $n = 10^3$ & $n = 10^4$ & $n = 10^5$\\ \hline\hline
	 & static & 66 $~$ & 213 &  677 * \\ \hhline{|~|-|-|-|-|}
%
   & $10\%$ & 70 & 224 $~$ &  714  $~$ \\ \hhline{|~|-|-|-|-|}
   $99\%$ & $30\%$ & 79 & 255 $~$  & 809  $~$  \\ \hhline{|~|-|-|-|-|}
   & $60\%$ & 105 & 337  $~$  &  1071 $~$ \\ \hhline{|~|-|-|-|-|}
   & $80\%$ & 143 & 478 $~$ &  1516 $~$  \\ \hline\hline
   
   & static & 80 * & 260 &  828 * \\ \hhline{|~|-|-|-|-|}
   
   & $10\%$ & 85 & 274 $~$ & 873 * \\ \hhline{|~|-|-|-|-|}
   $99.9\%$ & $30\%$ & 96 & 311  $~$ & 990 *\\ \hhline{|~|-|-|-|-|}
   & $60\%$ & 128 & 413 * &  1311 $~$ \\ \hhline{|~|-|-|-|-|}
   & $80\%$ & 182 & 584  $~$ & 1855  $~$ \\ \hline
  \end{tabular}
 }
 \caption{The core size depending on the system sizes and the churn rate.}
 \label{fig:tabular}
\end{figure} 

Figure~\ref{fig:tabular} focuses on  the core size that is required in a static
system and in a dynamic system (according to various values of the ratio $C$).  
The static system implies that no nodes leave or join the system while
the dynamic system contains nodes that join and leave the system depending on
several churn values.
For the sake of clarity we omit values of $\delta$ and 
simply present $C$ taking several values from $10\%$ to $80\%$.
The analysis of the results depicted in the figure leads to two interesting 
observations.

First, when $\delta$ is big enough for $10\%$ of the system 
nodes to be replaced,  then the core size required is
amazingly close to the static case 
(873 versus 828 when $n=10^5$ and the probability is $0.999$).
Moreover, $q$ has to be equal to $990$ only when $C$ increases up to $30\%$. 
Second, even when $\delta$ is sufficiently large to let $80\%$ 
of the system nodes be replaced, the minimal number of nodes
to probe remains low with respect to the system size.  For instance, 
if $\delta$ is sufficiently large to let $6,000$ nodes be
replaced in a system with $10,000$ nodes, then only 413 nodes  must be 
randomly probed to obtain an intersection with probability $p=0.999$.

To conclude, these results clearly show that a critical data
in a highly dynamic system can persist in a scalable way: 
even though the delay between core re-establishments is reasonably 
large while the  size  of the  core remains  relatively low. 


\section{Conclusion}
\label{sec:discussion}

Maintenance of critical data in large-scale dynamic systems where nodes 
may join and leave dynamically is a critical issue. In this paper, 
we define the notion of persistent core of nodes that can
maintain such critical data  with a high probability regardless of 
the structure of the underlying P2P network.  
More specifically, we relate
the parameters that can be tuned to achieve a high probability of 
defining a core, namely the size of the core, the
frequency at which it has to be re-established, and the churn rate 
of the system.

Our results provide application designers with a set of guidelines 
to tune the system
parameters depending on the expected guarantees and the churn rate variation.
An interesting outcome of this paper is
to 
show that slightly increasing the size of the core result in a significant 
probability increase of the guarantee. 

This work opens up a number of very interesting research directions. 
An interesting question
is related to the design and evaluation of efficient probing protocols, defining such a core in the system
applicable to a large spectrum of peer to peer overlay networks.  
Monitoring the system 
in order to estimate the churn rate is another interesting issue.

\bibliographystyle{abbrv}
\bibliography{P2P}

\end{document}